\documentstyle[12pt,epsf]{article}
 \newcommand{\insertplot}[5]{\begin{figure}
 \hfill\hbox to 0.05in{\vbox to #5in{\vfill
 \inputplot{#1}{#4}{#5}}\hfill}
 \hfill\vspace{-.1in}
 \caption{#2}\label{#3}
 \end{figure}}
 \newcommand{\inputplot}[3]{
 \special{ps: plotfile #1}
}

\begin{document}
 
\title{Non-Abelian Einstein-Born-Infeld Black Holes}
\vspace{1.5truecm}
\author{
{\bf Marion Wirschins, Abha Sood and Jutta Kunz}\\
Fachbereich Physik, Universit\"at Oldenburg, Postfach 2503\\
D-26111 Oldenburg, Germany}

\date{\today}

\maketitle
\vspace{1.0truecm}

\begin{abstract}
We construct regular and black hole solutions in
SU(2) Einstein-Born-Infeld theory.
These solutions have many features in common with the corresponding
SU(2) Einstein-Yang-Mills solutions.
In particular, sequences of neutral non-abelian solutions
tend to magnetically charged limiting solutions,
related to embedded abelian solutions.
Thermodynamic properties of the black hole solutions are addressed.
\end{abstract}

\vfill
\noindent {Preprint hep-th/0004130} \hfill\break
\vfill\eject

\section{Introduction}

Non-linear electrodynamics was proposed in the thirties
to remove singularities associated with charged pointlike
particles \cite{bi}.
More recently such non-linear theories were considered
in order to remove singularites associated with charged black holes
\cite{oli,bhsing,bhnosing}.

Among the non-linear theories of electrodynamics
Born-Infeld (BI) theory \cite{bi} is distinguished,
since BI type actions arise in many different contexts
in superstring theory \cite{bisuper1,bisuper2}.
The non-abelian generalization of BI theory 
yields an ambiguity in defining the Lagragian from
the Lie-algebra valued fields. 
While the superstring context favors a symmetrized trace \cite{sym}, 
the resulting Lagranian is
so far only known in a perturbative expansion \cite{symp}.
However, the ordinary trace structure has also been suggested \cite{tro}.

Motivated by this strong renewed interest 
in BI and non-abelian BI theory,
recently non-perturbative non-abelian BI solutions were studied,
both in flat and in curved space.
In particular, non-abelian BI monopoles and dyons exist 
in flat space \cite{gpss}
as well as in curved space \cite{ebih} together with dyonic BI black holes.

Likewise one expects regular and black hole solutions
in pure SU(2) Einstein-Born-Infeld (EBI) theory,
representing the BI generalizations of the regular Bartnik-McKinnon
solutions \cite{bm} and their hairy black hole counterparts
\cite{bh}.
This expectation is further nurtured by the observation,
that SU(2) BI theory possesses even in flat space 
a sequence of regular solutions \cite{bigal}.

Here we construct these SU(2) EBI
regular and black hole solutions,
employing the ordinary trace as in \cite{bigal}.
The set of EBI equations depends essentially on one parameter, $\gamma$,
composed of the coupling constants of the theory.
The solutions are labelled by the node number $n$ of the gauge field 
function.
We construct solutions up to node number $n=5$ and
discuss the limiting solutions,
obtained for $n \rightarrow \infty$,
for various values of the horizon radius $x_{\rm H}$ and 
the parameter $\gamma$.
We also address thermodynamic properties of the black hole solutions.

\section{SU(2) EBI Equations of Motion}

We consider the SU(2) EBI action
\begin{equation}
S=S_G+S_M=\int L_G \sqrt{-g} d^4x + \int L_M \sqrt{-g} d^4x
\ \label{action}  \end{equation}
with
\begin{equation}
L_G=\frac{1}{16\pi G}R \ , \ \ \
L_M=\beta^2 \: \left(1-\sqrt{1+\frac{1}{2 \beta^2} 
F^a_{\mu\nu} F^{a\mu\nu}-
\frac{1}{16 \beta^4} (F^a_{\mu\nu} \tilde F^{a\mu\nu}})^2\right)
\ , \label{lagm} \end{equation}
field strength tensor 
\begin{equation}
F^a_{\mu\nu}= \partial_\mu A^a_\nu - \partial_\nu A^a_\mu
            + e \epsilon^{abc} A^b_\mu A^c_\nu 
\ , \label{fmunu} \end{equation}
gauge coupling constant $e$, gravitational constant $G$
and BI parameter $\beta$.

To construct static spherically symmetric 
regular and black hole solutions
we employ Schwarz\-schild-like coordinates and adopt
the spherically symmetric metric
\begin{equation}
ds^2=g_{\mu\nu}dx^\mu dx^\nu=
  -{A}^2{N} dt^2 + {N}^{-1} dr^2 
  + r^2 (d\theta^2 + \sin^2\theta d\phi^2)
\ , \label{metric} \end{equation}
with the metric functions ${A}(r)$ and 
\begin{equation}
{N}(r)=1-\frac{2m(r)}{r}
\ . \label{N} \end{equation}
The static spherically symmetric and purely magnetic Ansatz
for the gauge field $A_{\mu}$ is
\begin{equation}
 A^a_0 = 0 \ , \ \ \ A^a_i = \epsilon_{aik} r^k \frac{1-w(r)}{er^2}
 \ . \label{amu} \end{equation}
For purely magnetic configurations $F_{\mu\nu} \tilde F^{\mu\nu}=0$.

With the ansatz (\ref{metric})-(\ref{amu})
we find the static action
\begin{eqnarray} 
S \,=\,e \int dr  \frac{A}{2} \left[ 1+N \left( 1 + 2r \left( \frac{A^{'}}{A}  
+\frac{N{'}}{2N}\right)\right)\right]
- \alpha^2 \beta^2 r^2 A \left[1 - \sqrt{1+\frac{2 N w{'}^2}{\beta^2 e^2 r^2}+
\frac{(1-w^2)^2}{\beta^2 e^2 r^4}}\right]
\ , \label{actst} \end{eqnarray}
where $\alpha^2 = 4 \pi G$.

Introducing dimensionless coordinates and a dimensionless mass function
\begin{equation}
 x = \sqrt{e\beta} r \ , \ \ \
\mu = \sqrt{e\beta} m 
 \   \label{dimless} \end{equation}
as well as the dimensionless parameter
\begin{equation}
\gamma = \frac{\alpha^2 \beta}{e}
\ , \label{gamma} \end{equation}
we obtain the set of equations of motion
\begin{eqnarray}
\mu^{'} & = & - \gamma x^2   
\left[1 - \sqrt{1+\frac{2 N w{'}^2}{x^2}+
\frac{(1-w^2)^2}{x^4}}\right]
\ , \label{sunm1} \\
\frac{ A^{'}}{ A} & = &\gamma \frac{2  w{'}^2}
{x \sqrt{1+\frac{2 N w{'}^2}{x^2}+
\frac{(1-w^2)^2}{x^4}}} 
\ , \label{suna1} \\
\left( A N \frac{w^{'}}{\sqrt{1+\frac{2 N w{'}^2}{ x^2}+
\frac{(1-w^2)^2}{x^4}}}\right)^{'} & = &
\frac{  A w (w^2 - 1)}{x^2 \sqrt{1+\frac{2 N w{'}^2}{x^2}+
\frac{(1-w^2)^2}{x^4}}}
\ . \label{sunu1} \end{eqnarray}
In eq.~(\ref{sunu1}) the metric function $A$ can be eliminated by means of
eq.~(\ref{suna1}).

We consider only asymptotically flat solutions, where the metric functions
$A$ and $\mu$ both approach a constant at infinity. Here we choose
\begin{equation}
A(\infty) = 1 
\ . \end{equation}
For magnetically neutral solutions the gauge field configuration 
approach a vacuum configuration at infinity
\begin{equation}
w(\infty)=\pm 1
\ . \label{bc0} \end{equation}
Globally regular solutions satisfy at the origin
the boundary conditions
\begin{equation}
\mu(0) = 0 \ , \ \ \ w(0)= 1
\ , \label{bc1} \end{equation}
whereas black hole solutions with a regular event horizon at $x_{\rm H}$
satisfy there
\begin{equation}
\mu(x_{\rm H})= \frac{x_{\rm H}}{2} \ , \ \ \
\left. N'w' \,\right|_{x_{\rm H}} = 
\left. \, \frac{w \left(w^2 -1\right)}{x^2} \right|_{x_{\rm H}}
\ . \label{bc2} \end{equation}

\section{Embedded Abelian BI Solutions}

Before discussing the non-abelian BI solutions,
let us briefly recall the embedded abelian BI solutions \cite{infeld,oli,rash}.
With constant functions $A$ and $w$,
\begin{equation}
A=1 \ , \ \ \ w=0
\ , \label{abel1} \end{equation}
they carry one unit of magnetic charge,
and their mass function satisfies
\begin{equation}
\mu' = \gamma \left( \sqrt{x^4+1} - x^2 \right)
\ , \label{abel2} \end{equation}
or upon integration
\begin{equation}
\mu(x) =\mu(0) + \gamma \int_0^x \left( \sqrt{{x'}^4+1} - {x'}^2 \right) dx'
\ . \label{abel3} \end{equation}
Their mass $\mu_\infty$ is thus given by
\begin{equation}
\mu_\infty =\mu(0) + \gamma \frac{\pi^{3/2}}{3 \Gamma^2 (3/4)}
\ . \label{abel4} \end{equation}

The solutions are classified according to the integration constant
$\mu(0)$ \cite{rash}.
For $\mu(0)>0$ black hole solutions
with one non-degenerate horizon $x_{\rm H}$
are obtained.
For $\mu(0)=0$ black hole solutions 
with one non-degenerate horizon $x_{\rm H}$ 
are obtained for $\gamma > 1/2$,
otherwise the solutions possess no horizon.
For $\mu(0)<0$ black hole solutions with two non-degenerate horizons,
extremal black hole solutions with one degenerate horizon
or solutions with no horizons are obtained,
similarly to the Reissner-Nordstr\o m (RN) case.

For black hole solutions with event horizon at
$ x_{\rm H}= 2 \mu(x_{\rm H}) $ 
the integration constant $\mu(0)$ is given by
\begin{equation}
\mu(0) = \frac{x_{\rm H}}{2} -
 \gamma \int_0^{x_{\rm H}} \left( \sqrt{x^4+1} - x^2 \right) dx 
\ . \label{abel8} \end{equation}
Since extremal black hole solutions satisfy
\begin{equation}
\mu'(x_{\rm H}) = \frac{\mu(x_{\rm H})}{x_{\rm H}} = \frac{1}{2}
\ , \label{abel6} \end{equation}
eq.~(\ref{abel2})
yields for the degenerate horizon of extremal black holes
\begin{equation}
x^{\rm ex}_{\rm H} = \sqrt{ \gamma - \frac{1}{4\gamma} }
\ . \label{abel7} \end{equation}
Thus we find a critical value $\gamma_{\rm cr}=1/2$, where
$x^{\rm ex}_{\rm H,cr} = 0$ and
$\mu_{\rm cr}(0) = 0$.
Finally, the Hawking temperature $T$ of the black holes
is given by \cite{rash}
\begin{equation}
 T = \frac{1}{4 \pi x_{\rm H}}
 \left[ 1 - 2 \gamma \left( \sqrt{x_{\rm H}^4 +1} - x_{\rm H}^2
 \right) \right]
\ . \label{temp} \end{equation}

\section{Regular Solutions}

Recently, a sequence of non-abelian regular BI solutions was found
in flat space by Gal'tsov and Kerner \cite{bigal},
labelled by the node number $n$ of the gauge field function $w$.
We now consider this sequence of BI solutions in the presence
of gravity, i.e.~for finite $\gamma$ (and thus finite $\alpha$),
and compare the non-abelian regular BI solutions
to the regular Einstein-Yang-Mills (EYM) solutions
\cite{bm}.

With increasing $\gamma$ the regular BI solutions evolve smoothly from
the corresponding flat space BI solutions. 
Let us first consider
the sequence of regular BI solutions for $\gamma_{\rm cr}$.
In Fig.~1 we show the gauge field function $w$ of
the regular BI solutions with node numbers $n=1$, 3 and 5.

The metric function $N$ of these BI solutions
is shown in Fig.~2
together with the metric function $N$ of the 
extremal abelian BI solution with $x^{\rm ex}_{\rm H}=0$ and 
one unit of magnetic charge.
Clearly, with increasing $n$
the metric function of the non-abelian regular BI solutions
tends to the metric function of the extremal abelian solution.

In Fig.~3 we show the charge function $P(x)$
\begin{equation}
P^2(x) = \frac{2 x}{\gamma} \left( \mu_\infty - \mu(x) \right)
\ , \label{charge} \end{equation}
obtained from an expansion of the abelian mass function
for general charge $P$,
for these regular BI solutions
together with the constant charge $P=1$ of the 
extremal abelian BI solution with $x^{\rm ex}_{\rm H}=0$.

Furthermore, for $\gamma = \gamma_{\rm cr}$
the masses of the non-abelian regular BI solutions
converge exponentially to the mass of the extremal abelian BI solution 
with $x^{\rm ex}_{\rm H}=0$ and unit magnetic charge,
which represents the limiting solution of this sequence.

Let us now consider $\gamma \ne \gamma_{\rm cr}$.
For $\gamma < \gamma_{\rm cr}$ 
the sequence of non-abelian regular BI solutions
tends to an abelian BI solution without horizon and unit
magnetic charge, and with integration constant $m(0)=0$.

For $\gamma > \gamma_{\rm cr}$ we must distinguish
two spatial regions,
$x < x^{\rm ex}_{\rm H}$ (given in eq.~(\ref{abel7})),
and $x > x^{\rm ex}_{\rm H}$.
Only in the region $x > x^{\rm ex}_{\rm H}$
the metric function $N$ of
the sequence of non-abelian regular BI solutions
tends to the metric function of 
the extremal abelian BI black hole solution with horizon
$x^{\rm ex}_{\rm H}$ and unit magnetic charge.
For $x < x^{\rm ex}_{\rm H}$
it tends to a non-singular limiting function.
This is demonstrated in Fig.~4, where the metric function $N$ is
shown for $n=5$ and $\gamma=0.01$, $\gamma=1$, and $\gamma=100$,
together with the metric function $N$ of the
corresponding limiting abelian BI solutions.

Thus the non-abelian regular BI solutions are very
similar to their EYM counterparts.
In particular, there with increasing node number $n$ the sequence 
of neutral $SU(2)$ EYM solutions also tends to a limiting
charged solution, which
for radius $x \ge 1$ is the 
extremal embedded RN solution with magnetic charge $P=1$ \cite{lim,kks3}.

\section{BI Black Hole Solutions}

Imposing the boundary conditions eq.~(\ref{bc2}),
leads to non-abelian BI black hole solutions with regular horizon.
Like their non-abelian EYM counterparts,
non-abelian BI black hole solutions exist for
arbitrary value of the horizon radius $x_{\rm H}$.
In both cases the sequences of neutral non-abelian BI black hole solutions
tend to limiting solutions with unit magnetic charge.

For $\gamma < \gamma_{\rm cr}$ 
the limiting solution of the non-abelian BI black hole solutions
is always the abelian BI black hole solution with the same horizon.
The same holds true for $\gamma = \gamma_{\rm cr}$.

For $\gamma > \gamma_{\rm cr}$ and
$x_{\rm H} < x^{\rm ex}_{\rm H}$
the limiting solution is the 
extremal abelian BI black hole solution with horizon $x^{\rm ex}_{\rm H}$

only in the region $x > x^{\rm ex}_{\rm H}$,
but differs from it in the region $x_{\rm H} < x < x^{\rm ex}_{\rm H}$.
For $x_{\rm H} > x^{\rm ex}_{\rm H}$
the limiting solution is always the
abelian BI black hole solution with the same horizon.
This is demonstrated in Fig.~5 for $\gamma=1$ and
horizon radii $x_{\rm H}=0.1$, 0.2, 0.5, and 10.

The temperature of the non-abelian BI black holes is obtained from \cite{T}
\begin{equation}
T = \frac{1}{4 \pi} A N'
\ . \label{temp2} \end{equation}
In Fig.~6 we show the inverse temperature of the non-abelian BI black hole
solutions with node numbers $n=1$, 3 and 5
as a function of their mass for $\gamma=0.01$ and $\gamma=1$.
Also shown is the inverse temperature of the corresponding limiting
abelian BI solutions.
Again, with increasing $n$ rapid convergence towards the limiting
values is observed, analogously to the EYM 
and EYM-dilaton case \cite{dil}.

Further details will be given elsewhere \cite{longer}.

\section{Conclusions}

We have constructed sequences of regular and black hole
solutions in SU(2) EBI theory. The solutions are labelled by the node number
$n$. With increasing node number
these sequences of non-abelian neutral solutions tend to 
limiting solutions, corresponding (at least in the
outer spatial region) to abelian BI solutions with unit magnetic charge.
These features are similar to those observed for
non-abelian EYM and EYM-dilaton regular and
black hole solutions \cite{bh,lim,dil,kks3}.
This similarity is also observed for the Hawking temperature.

By generalizing the framework of isolated horizons to non-abelian
gauge theories \cite{iso},
recently new results were obtained for EYM black hole solutions.
In particular nontrivial relations between the masses of
EYM black holes and regular EYM solutions were found,
and a general testing bed for the instability of non-abelian
black holes was suggested \cite{iso}.
Application of such considerations to non-abelian BI black holes 
appears to be interesting.

\newpage

\begin{figure}
\centering
\epsfysize=11cm
\mbox{\epsffile{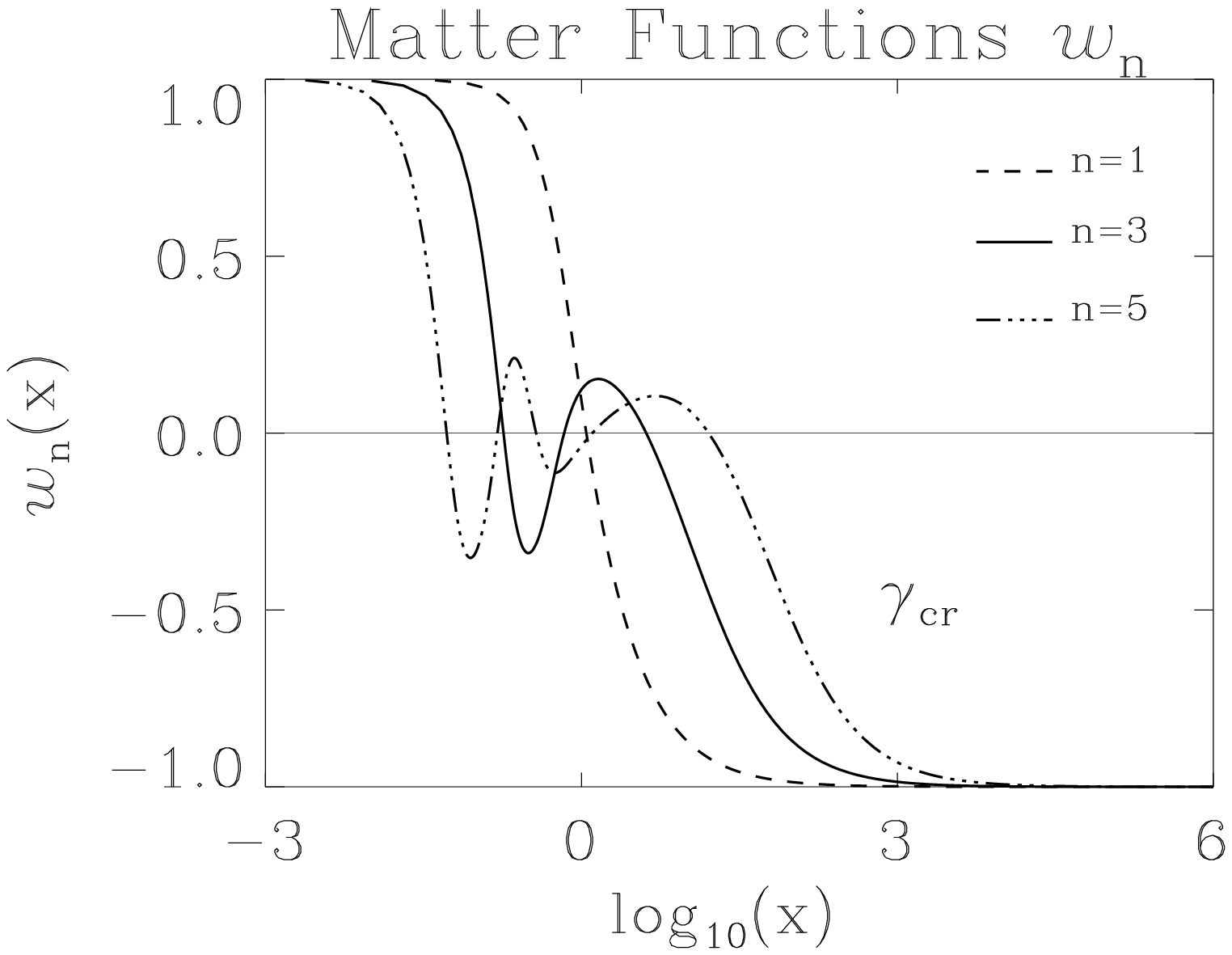}}
\caption{
The gauge field function $w$ is shown as a function of the dimensionless
coordinate $x$ for the regular BI solutions with node numbers
$n=1$, 3, 5 for $\gamma=\gamma_{\rm cr}$.
}
\end{figure}

\newpage

\begin{figure}
\centering
\epsfysize=11cm
\mbox{\epsffile{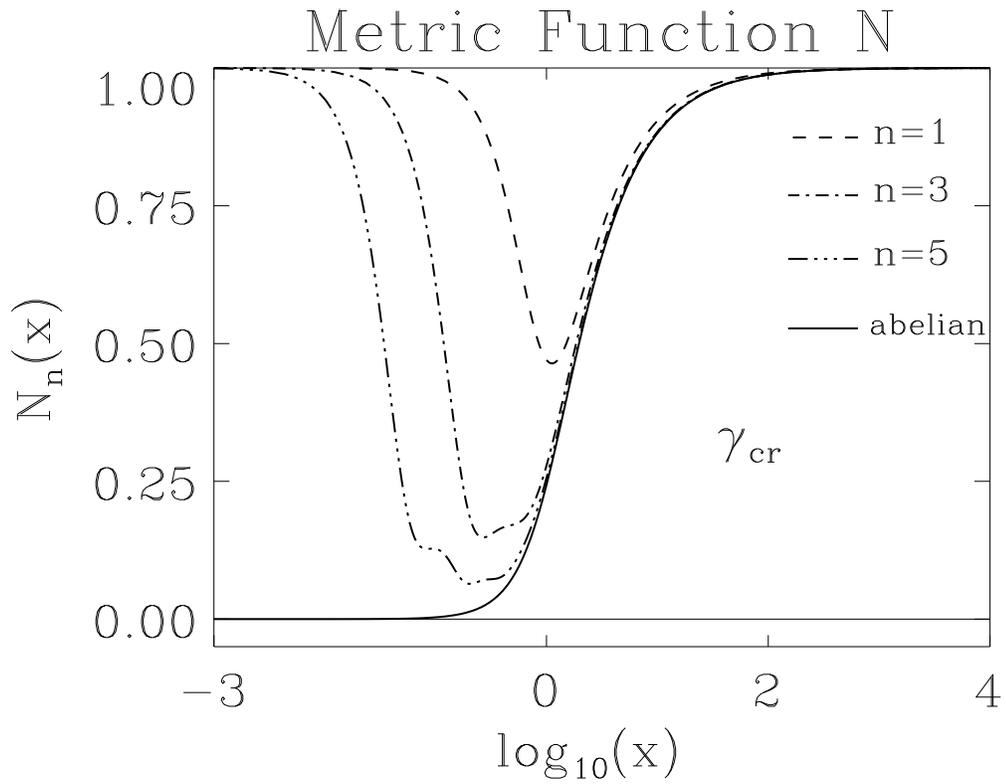}}
\caption{
The metric function $N$ is shown as a function of the dimensionless
coordinate $x$ for the regular BI solutions with node numbers
$n=1$, 3, 5 for $\gamma=\gamma_{\rm cr}$.
Also shown is the metric function $N$ for the
extremal abelian BI solution with $x^{\rm ex}_{\rm H}=0$ and 
unit magnetic charge.
}
\end{figure}

\newpage

\begin{figure}
\centering
\epsfysize=11cm
\mbox{\epsffile{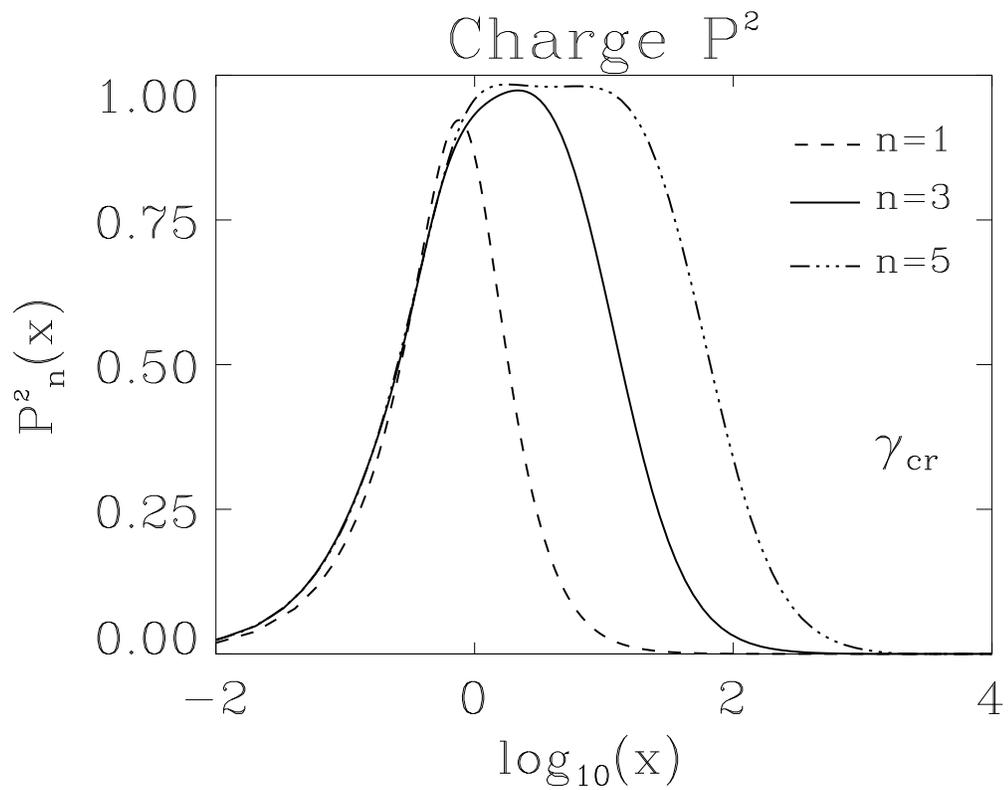}}
\caption{
The charge function $P^2$ is shown as a function of the dimensionless
coordinate $x$ for the regular BI solutions with node numbers
$n=1$, 3, 5 for $\gamma=\gamma_{\rm cr}$.
Also shown is the constant function $P=1$ of the abelian BI solution.
}
\end{figure}

\newpage

\begin{figure}
\centering
\epsfysize=11cm
\mbox{\epsffile{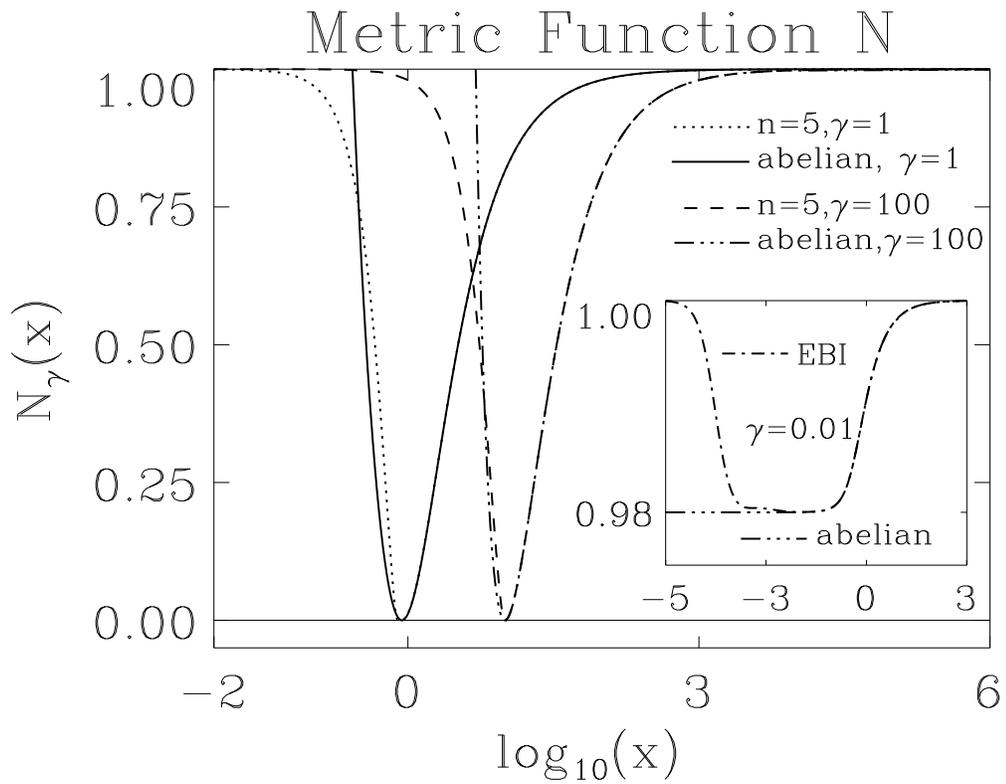}}
\caption{
The metric function $N$ is shown as a function of the dimensionless
coordinate $x$ for the regular BI solutions with node number
$n=5$ for $\gamma=0.01$, $\gamma=1$, and $\gamma=100$.
Also shown are the metric functions $N$ of the
corresponding limiting abelian BI solutions.
}
\end{figure}

\newpage

\begin{figure}
\centering
\epsfysize=11cm
\mbox{\epsffile{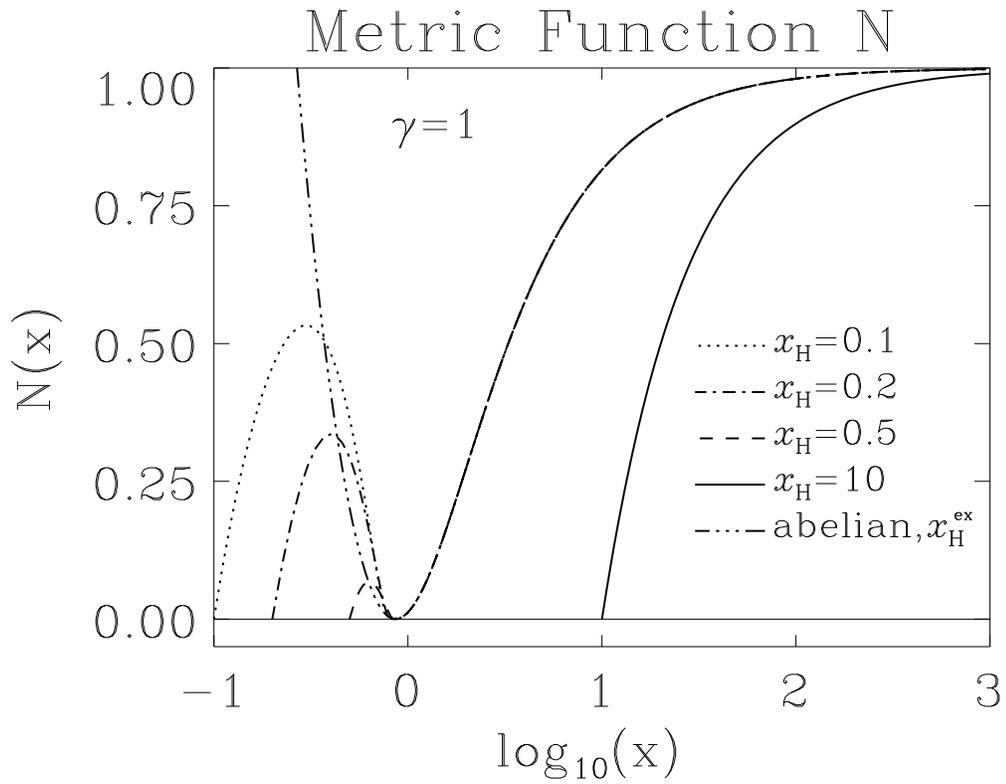}}
\caption{
The metric function $N$ is shown as a function of the dimensionless
coordinate $x$ for the BI black hole solutions with node number
$n=5$ for $\gamma=1$ and
horizon radii $x_{\rm H}=0.1$, 0.2, 0.5, and 10.
Also shown are the metric functions $N$ of the
corresponding limiting abelian BI solutions.
}
\end{figure}

\newpage

\begin{figure}
\centering
\epsfysize=11cm
\mbox{\epsffile{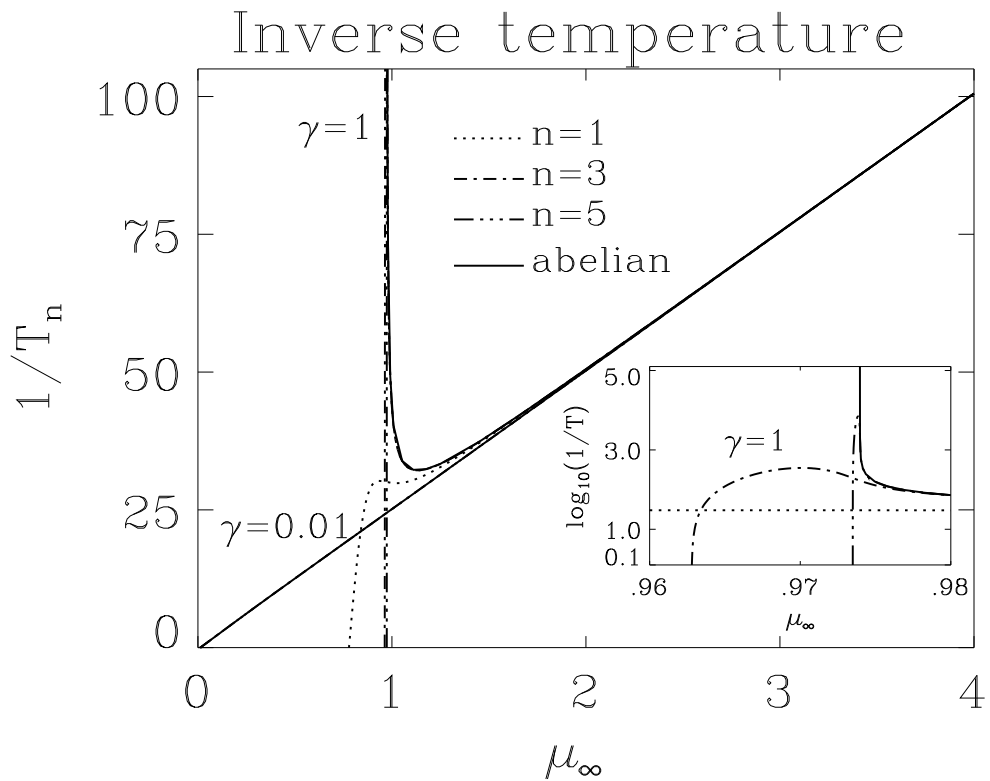}}
\caption{
The inverse temperature $1/T$ is shown as a function of the dimensionless
mass $\mu$ for the non-abelian BI black hole
solutions with node numbers $n=1$, 3 and 5
for $\gamma=0.01$ and $\gamma=1$.
Also shown is the inverse temperature for the corresponding limiting
abelian BI solutions.
(The $\gamma=0.01$ curves cannot be graphically resolved in the figure.)
}
\end{figure}

\end{document}